\documentclass[preprint,1p]{elsarticle}
\biboptions{sort&compress}

\usepackage{hyperref}

\journal{XXXXX}

\usepackage {amsmath}
\usepackage{amssymb}
\usepackage{graphicx}
\usepackage{subfigure}
\usepackage{color}
\usepackage{cases}
\usepackage{verbatim}
\usepackage{geometry}
\begin{document}

\begin{frontmatter}

\title{Generalized quantum teleportation of shared quantum secret with quantum walks}{Title for citation}

\author[a,b]{Hengji Li}

\author[b,c]{Jian Li\corref{cor1}}
\ead{lijian@bupt.edu.cn}

\author[c,d]{Xiubo  Chen}

\cortext[cor1]{Corresponding author}

\address[a]{Information Security Center, State Key Laboratory Networking and Switching Technology, Beijing University of Posts Telecommunications, Beijing 100876, China}
\address[b]{School of Artificial Intelligence, Beijing University of Posts Telecommunications, Beijing 100876, China}
\address[c]{Center for Quantum Information Research, ZaoZhuang University, ZaoZhuang Shandong 277160, China}
\address[d]{GuiZhou University, Guizhou Provincial Key Laboratory of Public Big Data, Guizhou Guiyang, 550025, China}

\begin{abstract}
Very recently, Lee {\it et al.} proposed the first secure quantum teleporation protocol, where quantum information shared by an arbitrary number of  senders can be transferred to another arbitrary number of receivers. Here, by introducing  quantum walks, a  novel secure $(n,m)$  quantum teleportation of shared quantum secret between $n$ senders and $m$ receivers is presented. Firstly, two kinds of  $(n,2)$ teleportation schemes are  proposed by $n$-walker quantum walks on the line, the first walker of which is driven by three coins, respectively, based on two kinds of coin operators: the homogeneous coins and the position-dependent coins. Secondly, by increasing the amount of the coins of the first walker to $m+1$, the previous $(n,2)$ scheme can be generalized to $(n,m)$ teleportation scheme. Then, we give the proof of the information security of our proposed scheme, in which  neither any single nor subparties of senders and receivers can fully access the secret quantum information. Moreover, the projective measurements are needed, instead of the joint Bell measurements that are necessary in Lee {\it et al.}'s protocol. Our work can also be extended further to QWs on the cycle. This work  provides an additional relevant instance of the richness of quantum walks for quantum information processing tasks and thus opens the wider application purpose  of quantum walks.
\end{abstract}


\end{frontmatter}

\section{Introduction}
\label{intro}

Quantum teleportation, first proposed by Bennett {\it et al.}\cite{Bennett1993Teleporting}, allows us  to transfer an unknown quantum state from the sender  to the receiver  without transferring the quantum particle.  Nowadays, it is a fundamental building block of formal quantum information theory\cite{nielsen2002quantum} and  quantum technology, such as quantum computation\cite{raussendorf2001one}, quantum network\cite{Hayashi2007Prior}, and quantum repeaters\cite{briegel1998quantum}.

The original protocol  transfers quantum information from one sender to one receiver, however, to realize the versatile quantum networks, it has been extended to the protocol that split quantum information from one sender to multiple receivers \cite{karlsson1998quantum,cleve1999share,hillery1999quantum,muralidharan2008quantum,dou2018secure}, in which no single receiver can fully access the information unless collaborated by all other receivers. Recently, Lee {\it et al.} proposed the first  teleportation protocol of shared quantum secret state  among  multiple senders  and receivers, in which a share quantum information is directly transferred to others among multiple parties without concentrating the information in the location of single or subparties \cite{lee2020quantum}. Following  Lee {\it et al.}'s work, by introducing quantum walks, we devote to presenting a novel kind of secure quantum teleportation scheme of shared quantum secret.

Quantum walks(QWs), the quantum mechanical analogs of classical random walks is a relatively hot research topic \cite{Aharonov1993Quantum,farhi1998quantum}. Two models of QWs have been suggested: discrete QWs (also call coined QWs)  \cite{Aharonov1993Quantum} and continuous QWs \cite{farhi1998quantum}. In coined QWs, an additional quantum system: a coin is introduced, which might produce plenty of interesting properties. Here, we focus on coined QWs.  Different classes of coined QWs have been explored in the past decades, such as QWs with many coins \cite{Brun2003Quantum,venegas2005quantum,liu2009one,liu2012asymptotic,shang2019quantum,shang2019experimental,
Wang2017Generalized,li2019neww,chandrashekar2011parrondo's,
Rajendran2018Implementing,rajendran2018playing,janexperimental,chatterjee2020experimental}, multiple walkers \cite{Omar2006Quantum,Pathak2007Agarwal,
vstefavnak2011directional,rohde2011multi,berry2011two,Peng2012Two,rigovacca2016two,
wang2016repelling,rohde2013increasing,li2019new,childs2013universal,li2013discrete,yang2018quantum}, and  nonhomogeneous coin operators \cite{konno2013limit,zhang2014one,suzuki2016asymptotic,ahmad2020one,montero2014invariance,panahiyan2018controlling,Yal2015Qubit,montero2017quantum,
2017Perfect,kurzynski2013quantum,li2019implementation,schreiber2010photons}.  Very recently, coined QWs have found several applications ranging from  quantum computation\cite{Lovett2010Universal,childs2013universal,rhodes2019quantum} to quantum information \cite{2017Perfect,shang2019quantum,shang2019experimental,
Wang2017Generalized,li2019new,chatterjee2020experimental,li2019neww,
kurzynski2013quantum,li2019implementation,chandrashekar2011parrondo's,
Rajendran2018Implementing,rajendran2018playing,janexperimental,Innocenti2017Quantum,li2013discrete,yang2018quantum}. For example, via utilizing the model of QWs with many coins, some secure quantum communication protocols are proposed \cite{shang2019quantum,shang2019experimental,Wang2017Generalized} and via utilizing the model of QWs with nonhomogeneous coin operators, generalized quantum measurement are
realized \cite{kurzynski2013quantum,li2019implementation}.

As the coherent action of the step and coin operators could lead to entanglement between the position and the coin state, different kinds of modes of QWs would produce  various of entanglement resources depending on homogeneous and nonhomogeneous coin operators, which  might open exciting new possibilities for the realization of quantum information protocols. In this manuscript, via using the model of multi-walker quantum walks with multiple coins, a novel  $(n,m)$  quantum teleportation protocol of shared quantum secret state  between $n$ senders and $m$ receivers is presented for the first time.  Based on two different coin operators: the \emph{homogeneous coins} and the \emph{position-dependent coins}, two corresponding novel $(n,2)$ teleportation schemes are presented with $n$-walker QWs, the first walker of which is driven by three coins.  And then  by increasing the amount of the coins of the first walker to $m+1$,  the previous $(n,2)$ scheme can be generalized to the  $(n,m)$ teleportation scheme.

In our proposed scheme, neither any single nor subparties of senders and receivers can fully access the secret quantum information and it allows us to relay quantum information over a network in an efficient and distributed manner without requiring fully trusted central or intermediate nodes. In addition,  the prior entangled state is not necessarily prepared, as the shift operator can  generate  entanglement between the position and the coin state. Moreover, the projective measurements are needed, instead of multiple times of standard  distributed Bell-state measurements that are necessary in Lee {\it et al.}'s protocol \cite{lee2020quantum}. Our results provide an additional relevant instance of the richness of QWs for quantum information processing tasks, notwithstanding their simplicity. This work thus opens the wider application purpose  of quantum walks and shows a new route to the realization of secure distributed quantum communications in quantum network.

The paper is structured as follows. First the  preliminaries are provided about the knowledge of QWs.  In Sec.\ref{qubit}, two novel kinds of  $(n,2)$ and $(n,m)$ schemes of shared quantum secret state are presented, respectively, based on two kinds of coin operators: the homogeneous coins and the position-dependent coins.  Then, we prove the information security of our scheme and discuss our main results.  Finally, Sec.\ref{Conclusion} contains our conclusions.

\section{preliminaries}
\label{preliminaries}
\subsection{QWs on the line}
QWs on the line \cite{ambainis2001one} takes place in
the product space $\mathcal{H}=\mathcal{H}_p\otimes\mathcal{H}_c$, with $\mathcal{H}_p$ as the position Hilbert space  $\{\left|x\right\rangle |x \in Z \}$, and  $\mathcal{H}_c$ as the two-dimensional auxiliary ``coin" space  $\{\left|0\right\rangle,\left|1\right\rangle\}$. One step of the walk is defined  by  $\mathcal{U}=\mathcal{S}(\mathbb{I}\otimes \mathbb{C})$
where $\mathcal{S}$ is the conditional shift operator, $\mathbb{C}$ is  any unitary operator, and $\mathbb{I}$ is the identity operator. The shift operator is defined as
\begin{equation}
\mathcal{E}=\mathcal{S}\otimes\left|0\right\rangle\left\langle 0\right|+\mathcal{S}^{\dag}\otimes\left|1\right\rangle\left\langle 1\right|,
\end{equation}
with  $\mathcal{S}=\sum_{x}\left|x+1\right\rangle\left\langle x\right|$ and $\mathcal{S}^{\dag}=\sum_{x}\left|x-1\right\rangle\left\langle x\right|$. Consequently, after $t$ steps
of the walk the initial  state
$\left|\psi_0\right\rangle$ will become $\left|\psi_{t}\right\rangle=\mathcal{U}^{t}\left|\psi_{0}\right\rangle$ .

\subsection{QWs with multiple coins}
Brun {\it et al.} presented the model of QWs with many coins, which can be utilized as one possible route to classical behavior of coined QWs \cite{Brun2003Quantum}. In this model, one could replace the coin with a new quantum coin for each flip and after a time $t$, one would have accumulated $t$ coins, all of which are  entangled with the position of the particle. At each step, the particle moves in the direction dictated by the coin that is \emph{active} at that step, with the other coins remaining inert until it is their turn once again. Therefore, the unitary transformation that results from flipping the $m$-th coin is
\begin{equation}
\mathcal{E}_{m}=\mathcal{S}_{m}\otimes\left|0\right\rangle\left\langle 0\right|+\mathcal{S}_{m}^{\dag}\otimes\left|1\right\rangle\left\langle 1\right|,
\end{equation}
with $\mathcal{S}_{m}=\mathcal{S}$.  As the introduction of the new coins might produce some novel properties, the model has been studied widely \cite{venegas2005quantum,liu2009one,liu2012asymptotic} and been used as a tool to design some quantum communication protocols \cite{shang2019quantum,shang2019experimental,Wang2017Generalized,li2019new} and quantum Parrondo's game \cite{chandrashekar2011parrondo's,
Rajendran2018Implementing,rajendran2018playing,janexperimental}.

\subsection{QWs with multiple walkers}
With involving more walkers,  QWs with multiple walkers are extensively researched \cite{Omar2006Quantum,Pathak2007Agarwal,
vstefavnak2011directional,rohde2011multi,Peng2012Two,rigovacca2016two,
wang2016repelling,rohde2013increasing,
childs2013universal,berry2011two,li2019new,li2013discrete,yang2018quantum}, which unlock additional possibilities which are worth investigating, for instance, the two walkers can be also entangled while the initial coin states are entangled \cite{Omar2006Quantum}. Multi-walker QWs  could be  capable of universal quantum computation \cite{childs2013universal} and can be used as a tool in some fields such as graph isomorphism testing \cite{berry2011two}, constructing quantum Hash scheme \cite{li2013discrete}, and quantum network coding \cite{yang2018quantum}. QWs with $n$ walkers takes place in the Hilbert space
$\mathcal{H}=\otimes_{i=1}^{n}\mathcal{H}_i$ with $\mathcal{H}_i=\mathcal{H}_{p_i}\otimes
\mathcal{H}_{c_i}$ for walker $i$ and thus each step of  QWs will be given by $\mathcal{U}_{1,..,n}=\otimes_{i=1}^{n}\mathcal{U}_i$, where $\mathcal{U}_i$ is the unitary operator on the
$i$-th walker.
\subsection{QWs with nonhomogeneous coin operators}
In the early models, researchers devoted on the study on homogeneous coin operators\cite{ambainis2001one}, one of most studied which is Hadamard ($\mathbb{H}$) coin. Later, QWs with various of nonhomogeneous coin operators are  studied widely, such as site-dependent coins \cite{konno2013limit,zhang2014one,
suzuki2016asymptotic,ahmad2020one}, time-dependent coins \cite{montero2014invariance,panahiyan2018controlling} as well as  site- and time-dependent coins \cite{Yal2015Qubit,montero2017quantum,2017Perfect,kurzynski2013quantum,li2019implementation}, which make it possible to obtain the desired distribution.  In the homogeneous models, the coin operator is independent of the time and position, however, in the  nonhomogeneous models, it will change and the new unitary operator can be denoted as $\mathbb{C}(p,t)$ with $p$ and  $t$ respectively representing  the position of the walker and the steps of the evolution. The ability to operate with different coins and the ease of addressing individual position states opens exciting new possibilities for the realization of quantum information
protocols \cite{schreiber2010photons}, such as perfect state transfer \cite{2017Perfect} and generalized quantum measurement \cite{kurzynski2013quantum,li2019implementation}.

\section{Scheme of teleportation between multiple senders and receivers}\label{qubit}
Here, we succeed in teleporting shared quantum secret from the \emph{senders}, i.e., a group of $n$ parties to  the \emph{receivers}, i.e., another group of $m$ parties, by utilizing multi-walker QWs with multiple coins.  Firstly, two  novel $(n,2)$ teleportation schemes are presented with $n$-walker QWs, the first walker of which is driven by three coins, based on two different coin operators: the \emph{homogeneous coins} and the \emph{position-dependent coins}. Secondly,  the previous $(n,2)$ scheme can be generalized to the  $(n,m)$ teleportation scheme, by increasing the amount of the coins of the first walker to $m+1$. The receivers can reconstruct and share the secret by the appropriate joint work of local operations, and no participants can access the full secret quantum information during the whole process. Suppose that a quantum secret in $|\phi\rangle=\alpha|0\rangle_{L}+\beta|1\rangle_{L}$ with logical basis, $|0\rangle_{L}$ and  $|1\rangle_{L}$, is shared by separated $n$ parties in quantum network. The shared secret state  to be teleported  can be written as $|\Phi\rangle_{s}^{n}=\alpha\otimes_{i=1}^{n}|0\rangle_{s_{i}}+\beta \otimes_{i=1}^{n}|1\rangle_{s_{i}}$.

\subsection{Scheme of teleportation between $n$ senders and $2$ receivers}\label{n2}

Initially, $n$ senders hold $n$ particles $s_{p_i}$ and $s_{c_{i,1}}(i=1,2,\cdots,n)$, respectively, which are corresponding to the position  and the first coin of the $i$-th walker, respectively. The shared state  $|\Phi\rangle_{s}^{n}$ is encoded in $n$ particles $s_{c_{i,1}}$.
Similarly,  the receiver $\emph{\textbf{r}}_{1}$ and $\emph{\textbf{r}}_{2}$ have one particle $r_{c_{1,2}}$ and $r_{c_{1,3}}$, respectively, which are corresponding to the second and third coin of the
first walker.  Thus, the joint Hilbert space of the composite system is $\mathcal{H}_{1,2,\cdots,n}=\otimes_{i=1}^{n}\mathcal{H}_{p_i}
\otimes_{i=1}^{n}\mathcal{H}_{c_{i,1}}\otimes\mathcal{H}_{c_{1,2}}\otimes\mathcal{H}_{c_{1,3}}$.

Next, two steps are needed to accomplish the task. To begin with, we would run the first step QW of the $n$ walkers, in which the first coin $s_{c_{i,1}}$ is the \emph{active } coin. Consequently,  the composite unitary operator can be written as $\mathcal{U}^{(1)}=\mathcal{E}_{1}\mathcal{C}_{1}$, $\mathcal{C}_{1}$ defined by
\begin{subequations}\label{U1}
\begin{align}
&\mathcal{C}_{1}=\otimes_{i=1}^{n}(\mathbb{I}_{p_{i}}\otimes\mathbb{C}_{c_{i,1}})
(\mathbb{I}_{c_{1,2}}\otimes\mathbb{I}_{c_{1,3}}),\\
&\mathcal{E}_{1}=\otimes_{i=1}^{n}(\mathcal{S}_{p_{i}}\otimes|0\rangle_{c_{i,1}}\langle 0|+\mathcal{S}_{p_{i}}^{\dag}\otimes|1\rangle_{c_{i,1}}\langle 1|)(\mathbb{I}_{c_{1,2}}\otimes\mathbb{I}_{c_{1,3}}),
\end{align}
\end{subequations}
where $\mathbb{I}$  is the identity operator, $\mathbb{C}$ is the coin operator and $S$ is the conditional shift operator, and the subscript represents which particle the unitary operator are performed on, i.e., $\mathbb{C}_{c_{i,1}}$ is the coin operation on the first coin of walker $i$. Next, in the second step,  QWs are performed on the second ($r_{c_{i,2}}$) and third ($r_{c_{i,3}}$) coin of the first walker, and then the corresponding unitary operator can be written as $\mathcal{U}^{(2)}=\mathcal{E}_{3}\mathcal{C}_{3}\mathcal{E}_{2}\mathcal{C}_{2}$, defined by
\begin{subequations}\label{U2}
\begin{align}
\mathcal{C}_{2}=&\otimes_{i=1}^{n}(\mathbb{I}_{p_{i}}\otimes\mathbb{I}_{c_{i,1}})
\otimes\mathbb{C}_{c_{1,2}}\otimes\mathbb{I}_{c_{1,3}},\\
\mathcal{E}_{2}=&\mathcal{S}_{p_{1}}\!\!\otimes\mathbb{I}_{c_{1,1}}\!\!\otimes_{i=2}^{n}(\mathbb{I}_{p_{i}}\otimes\mathbb{I}_{c_{i,1}})
\otimes|0\rangle\!_{c_{1,2}}\!\langle0|\otimes\mathbb{I}_{c_{1,3}}\!+\!\mathcal{S}_{p_{1}}^{\dag}\otimes\mathbb{I}_{c_{1,1}}\otimes_{i=2}^{n}(\mathbb{I}_{p_{i}}\otimes\mathbb{I}_{c_{i,1}})
\otimes|1\rangle\!_{c_{1,2}}\!\langle1|\otimes\mathbb{I}_{c_{1,3}},\\
\mathcal{C}_{3}=&\otimes_{i=1}^{n}(\mathbb{I}_{p_{i}}\otimes\mathbb{I}_{c_{i,1}})
\otimes\mathbb{I}_{c_{1,2}}\otimes\mathbb{C}_{c_{1,3}},\\
\mathcal{E}_{3}=&\mathcal{S}_{p_{1}}\!\!\otimes\mathbb{I}_{c_{1,1}}\!\!\otimes_{i=2}^{n}(\mathbb{I}_{p_{i}}\otimes\mathbb{I}_{c_{i,1}})
\otimes\mathbb{I}_{c_{1,2}}\otimes|0\rangle\!_{c_{1,3}}\!\langle0|\!+\!\mathcal{S}_{p_{1}}^{\dag}\otimes\mathbb{I}_{c_{1,1}}\otimes_{i=2}^{n}(\mathbb{I}_{p_{i}}\otimes\mathbb{I}_{c_{i,1}})
\otimes\mathbb{I}_{c_{1,2}}\otimes|1\rangle\!_{c_{1,3}}\!\langle1|.
\end{align}
\end{subequations}

Then, after running two steps, every separate sender $\emph{\textbf{s}}_{i}$  performs two single-particle measurements on  $s_{p_i}$ and $s_{c_{i,1}}$ and denote the results as $p_{i}$ and $c_{i}$, respectively. Then the senders communicate the outcomes with the receivers. Conditioned on the measured results, the receivers can know the quantum state at their locations and can share the quantum secret $|\phi\rangle$.

Assume that the position state of each walker is initially $\left|0\right\rangle$ and the coin state in $r_{c_{1,2}}$ and $r_{c_{1,3}}$ is $\left|0\right\rangle$. Therefore,  the overall initial tensor state will be
\begin{equation}\label{initiaN}
\left|\Psi\right\rangle=\otimes_{i=1}^{n} \left|0\right\rangle_{p_{i}}\otimes(\alpha\otimes_{i=1}^{n}|0\rangle_{c_{i,1}}+\beta \otimes_{i=1}^{n}|1\rangle_{c_{i,1}})\otimes\left|0\right\rangle_{c_{1,2}}
\otimes\left|0\right\rangle_{c_{1,3}},
\end{equation}
For simplifying the expression, we will omit the subscripts in the following statement. In the case that we take $\mathbb{C}_{c_{i,1}}=\mathbb{I}$ as the coin operators in the first step without loss of generality,  the initial state (\ref{initiaN}) thus will evolve into
\begin{equation}\label{mid1}
\left|\Psi^{\prime}\right\rangle=\alpha\left|1\right\rangle^{\otimes n}\left|0\right\rangle^{\otimes n}\left|0\right\rangle\left|0\right\rangle
+\beta\left|-1\right\rangle^{\otimes n}\left|1\right\rangle^{\otimes n}\left|0\right\rangle\left|0\right\rangle,
\end{equation}

There are many types of  coin operators such as nonhomogeneous and site-dependent coin operators \cite{konno2013limit,zhang2014one,
suzuki2016asymptotic,ahmad2020one}, which will generate various of different entanglement resources. In order to accomplish the task, two kinds of coin operators, the\emph{ homogeneous coins} and the \emph{position-dependent coins}, are utilized to produce the entanglement state in the second step.

\textbf{\emph{Case 1: homogeneous coins.}} Taking $\mathbb{C}_{c_{1,2}}=\mathbb{C}_{c_{1,3}}=\mathbb{H}$ as the coin operators (the choice of $\mathbb{H}$ is for producing the maximum entanglement state), the state (\ref{mid1}) will evolve into
\begin{equation}\label{fline1}
\begin{split}
\left|\Psi^{\prime\prime}_{h}\right\rangle&=\alpha\left|3\right\rangle\left|1\right\rangle^{\otimes (n-1)}\left|0\right\rangle^{\otimes n}\left|0\right\rangle\left|0\right\rangle
+\alpha\left|1\right\rangle\left|1\right\rangle^{\otimes (n-1)}\left|0\right\rangle^{\otimes n}\left|0\right\rangle\left|1\right\rangle+\\
&\alpha\left|1\right\rangle\left|1\right\rangle^{\otimes (n-1)}\left|0\right\rangle^{\otimes n}\left|1\right\rangle\left|0\right\rangle
+\alpha\left|-1\right\rangle\left|1\right\rangle^{\otimes (n-1)}\left|0\right\rangle^{\otimes n}\left|1\right\rangle\left|1\right\rangle+\\
&\beta\left|1\right\rangle\left|-1\right\rangle^{\otimes (n-1)}\left|1\right\rangle^{\otimes n}\left|0\right\rangle\left|0\right\rangle+\beta\left|-1\right\rangle\left|-1\right\rangle^{\otimes (n-1)}\left|1\right\rangle^{\otimes n}\left|0\right\rangle\left|1\right\rangle+\\
&\beta\left|-1\right\rangle\left|-1\right\rangle^{\otimes (n-1)}\left|1\right\rangle^{\otimes n}\left|1\right\rangle\left|0\right\rangle+\beta\left|-3\right\rangle\left|-1\right\rangle^{\otimes (n-1)}\left|1\right\rangle^{\otimes n}\left|1\right\rangle\left|1\right\rangle,
\end{split}
\end{equation}
which is not normalized. For brevity, we will ignore the normalization of quantum states in the following description. Furthermore, for understanding the quantum state (\ref{fline1}) more intuitively, by adjusting the order of the terms we can rewrite the final state (\ref{fline1}) as
\begin{equation}\label{flineNR}
\begin{split}
\left|\Psi^{\prime\prime}_{h}\right\rangle&=\underbrace{\alpha\left|3\right\rangle\left|1\right\rangle^{\otimes (n-1)}\left|0\right\rangle^{\otimes n}\left|0\right\rangle\left|0\right\rangle
+\beta\left|-1\right\rangle\left|-1\right\rangle^{\otimes (n-1)}\left|1\right\rangle^{\otimes n}\left|0\right\rangle\left|1\right\rangle}+\\
&\underbrace{\alpha\left|-1\right\rangle\left|1\right\rangle^{\otimes (n-1)}\left|0\right\rangle^{\otimes n}\left|1\right\rangle\left|1\right\rangle+\beta\left|-1\right\rangle\left|-1\right\rangle^{\otimes (n-1)}\left|1\right\rangle^{\otimes n}\left|1\right\rangle\left|0\right\rangle}+\\
&\underbrace{\alpha\left|1\right\rangle\left|1\right\rangle^{\otimes (n-1)}\left|0\right\rangle^{\otimes n}\left|0\right\rangle\left|1\right\rangle+\beta\left|1\right\rangle\left|-1\right\rangle^{\otimes (n-1)}\left|1\right\rangle^{\otimes n}\left|0\right\rangle\left|0\right\rangle}+\\
&\underbrace{\alpha\left|1\right\rangle\left|1\right\rangle^{\otimes (n-1)}\left|0\right\rangle^{\otimes n}\left|1\right\rangle\left|0\right\rangle+\beta\left|-3\right\rangle\left|-1\right\rangle^{\otimes (n-1)}\left|1\right\rangle^{\otimes n}\left|1\right\rangle\left|1\right\rangle},
\end{split}
\end{equation}
where the two terms enclosed in parentheses  are the ultima quantum state obtained by the receivers. Then,  each sender $\textbf{\emph{s}}_{i}$ performs two single-particle measurements on $s_{p_{i}}$ and  $s_{c_{i,1}}$ with the corresponding basis
$$ \left\{
\begin{aligned}
\Lambda_{h} = &\{|\hspace{-3pt}-\hspace{-3pt}\tilde{3}\rangle,|\hspace{-3pt}-\hspace{-3pt}\tilde{1}\rangle,
|\tilde{1}\rangle,|\tilde{3}\rangle\},\,s_{p_{1}}, \\
\Theta_{h} =&\{|\hspace{-3pt}-\hspace{-3pt}\hat{1}\rangle,|\hat{1}\rangle\},\,s_{p_{k}},k=2,\cdots,n, \\
\Delta_{c} =& \{\left|-\right\rangle,\left|+\right\rangle\},\,s_{c_{i,1}},i=1,2,\cdots,n,
\end{aligned}
\right.
$$
with
$$
\left\{
\begin{aligned}
|\tilde{3}\rangle=&(\left|3\right\rangle+\left|\hspace{-1pt}-\hspace{-1pt}1\right\rangle)/\sqrt{2},
|\hspace{-2pt}-\hspace{-2pt}\tilde{1}\rangle=(\left|3\right\rangle-\left|\hspace{-1pt}-\hspace{-1pt}1\right\rangle)/\sqrt{2},
|\tilde{1}\rangle=(\left|1\right\rangle+\left|\hspace{-1pt}-\hspace{-1pt}3\right\rangle)/\sqrt{2},
|\hspace{-2pt}-\hspace{-2pt}\tilde{3}\rangle=(\left|1\right\rangle-\left|\hspace{-1pt}-\hspace{-1pt}3\right\rangle)/\sqrt{2},\\
|\hat{1}\rangle=&(\left|1\right\rangle+\left|\hspace{-1pt}-\hspace{-1pt}1\right\rangle)/\sqrt{2},
|\hspace{-2pt}-\hspace{-2pt}\hat{1}\rangle=(\left|1\right\rangle-\left|\hspace{-1pt}-\hspace{-1pt}1\right\rangle)/\sqrt{2},\\
|+\rangle=&(\left|1\right\rangle+\left|0\right\rangle)/\sqrt{2},
|-\rangle=(\left|1\right\rangle-\left|0\right\rangle)/\sqrt{2},\\
\end{aligned}
\right.
$$
Mark the measurements results 00, 01, 10 and 11 corresponding $|\tilde{3}\rangle$, $|\hspace{-2pt}-\hspace{-2pt}\tilde{1}\rangle$, $|\tilde{1}\rangle$ and $|\hspace{-2pt}-\hspace{-2pt}\tilde{1}\rangle$ for $s_{p_{1}}$. And mark  the results 1 and 0 corresponding to $|\hspace{-2pt}-\hspace{-2pt}\hat{1}\rangle$($|-\rangle$) and $|\hat{1}\rangle$($|+\rangle$) for $s_{p_{k}}$($s_{c_{i,1}}$) . Therefore, conditioned on the results, the final state obtained by the receivers can be written as
\begin{subnumcases}
{|\Phi_{h}\rangle_{r}^{2}=}\label{n21}
|\Phi_{h,s}\rangle_{r}^{2}: &$(-1)^{s}\alpha|00\rangle+\alpha|11\rangle+(-1)^{\omega_{h}}(\beta|01\rangle+\beta|10\rangle),p_{1}=0s$,\\\label{n22}
|\Phi_{h,t}\rangle_{r}^{2}: &$\alpha|01\rangle+\alpha|10\rangle+(-1)^{\omega_{h}}(\beta|00\rangle+(-1)^{t}\beta|11\rangle),p_{1}=1t$.
\end{subnumcases}
where $s,t \in\{0,1\}$, $\omega_{h}=(\sum_{k=2}^{n}p_{k}+\sum_{i=1}^{n}c_{i})\, \text{mod}\, 2$.
The initial state shared by $n$ receivers is $|\Phi\rangle_{s}^{n}=\alpha\otimes_{i=1}^{n}|0\rangle_{s_{i}}+\beta \otimes_{i=1}^{n}|1\rangle_{s_{i}}$ and taking $n=2$, the corresponding shared state will be  $|\Phi\rangle_{r}^{2}=|\Phi\rangle_{s}^{2}=\alpha|00\rangle+\beta|11\rangle$.
In Lee {\it et} al.'s protocol\cite{lee2020quantum}, the ultima state at the receivers' locations is  $|\Phi\rangle_{r}^{2}$ plus \emph{logical} Pauli operations and the receivers can carry out the \emph{local} Pauli operations to retrieve $|\Phi\rangle_{r}^{2}$. However, comparing the quantum state (\ref{n21}) and (\ref{n22}) with  the  shared quantum state $|\Phi\rangle_{r}^{2}$,  they look quite different intuitively. For clarify, rewrite $|\Phi\rangle_{r}^{2}$ as
\begin{equation}
|\Phi\rangle_{r}^{2}=|+\rangle|\phi\rangle+|-\rangle\sigma_{z}|\phi\rangle,
\end{equation}
Taking the result $p_{1}=0s$ as an example to analyze it. In this case, rewrite the final state  (\ref{n21}) under the measuring basis $\{\left|0\right\rangle,\left|1\right\rangle\}$  as
\begin{equation}\label{ini1}
(-1)^{s}\alpha|00\rangle+\alpha|11\rangle+(-1)^{\omega_{h}}(\beta|01\rangle+\beta|10\rangle)=
\left|0\right\rangle\sigma_{z}^{\omega_{h}+s}\left|\phi\right\rangle
+\left|1\right\rangle\sigma_{x}\sigma_{z}^{\omega_{h}}\left|\phi\right\rangle,
\end{equation}

Then $\textbf{\emph{r}}_{1}$ and $\textbf{\emph{r}}_{2}$ can respectively perform $\sigma_{z}^{\omega_{h}}$ and $\sigma_{z}^{\omega_{h}}\sigma_{x}$, which will make (\ref{ini1}) become
$|0\rangle\sigma_{x}\sigma_{z}^{s}|\phi\rangle+|1\rangle|\phi\rangle$.
Considering the  equation
$\left|0\right\rangle\sigma_{z}^{\omega_{h}+s}\left|\phi\right\rangle
+\left|1\right\rangle\sigma_{x}\sigma_{z}^{\omega_{h}}\left|\phi\right\rangle=\sigma_{z}^{\omega_{h}+s}\left|\phi\right\rangle
\left|0\right\rangle+\sigma_{x}\sigma_{z}^{\omega_{h}}\left|\phi\right\rangle\left|1\right\rangle$, it means that $\textbf{\emph{r}}_{1}$ and $\textbf{\emph{r}}_{2}$ can also perform respectively $\sigma_{z}^{\omega_{h}}\sigma_{x}$ and $\sigma_{z}^{\omega_{h}}$, which will similarly make (\ref{ini1}) be
$\sigma_{x}\sigma_{z}^{s}|\phi\rangle|0\rangle+|\phi\rangle|1\rangle$. It is obviously that
$|0\rangle\sigma_{x}\sigma_{z}^{s}|\phi\rangle+|1\rangle|\phi\rangle
=\sigma_{x}\sigma_{z}^{s}|\phi\rangle|0\rangle+|\phi\rangle|1\rangle$. Therefore, any of the receivers $\{\textbf{\emph{r}}_{1},\textbf{\emph{r}}_{2}\}$ performs the unitary operator $\sigma_{z}^{\omega_{h}}$, and the other one performs the unitary operator $\sigma_{z}^{\omega_{h}}\sigma_{x}$, to obtain the final quantum state at receivers' locations
\begin{equation}
|\Phi_{h,s}^{\prime}\rangle_{r}^{2}
=|0\rangle\sigma_{x}\sigma_{z}^{s}|\phi\rangle+|1\rangle|\phi\rangle,
\end{equation}
Similarly,  in the case that the result is $1t$, it can be deduced that the receivers $\{\textbf{\emph{r}}_{1},\textbf{\emph{r}}_{2}\}$ can also perform the same unitary operators as  the result is $0s$, however, it will yield
\begin{equation}
|\Phi_{h,t}^{\prime}\rangle_{r}^{2}
=|0\rangle|\phi\rangle+|1\rangle\sigma_{x}\sigma_{z}^{t}|\phi\rangle,
\end{equation}

To sum up,  after the QWs and the \emph{local} Pauli operators, eventually the initial shared quantum state $|\Phi\rangle_{s}^{n}$  becomes
\begin{equation}\label{phihr2}
|\Phi_{h}^{\prime}\rangle_{r}^{m}=\left\{
\begin{aligned}
&p_{1}=0s: |\Phi_{h,s}^{\prime}\rangle_{r}^{2},\\
&p_{1}=1t: |\Phi_{h,t}^{\prime}\rangle_{r}^{2}.
\end{aligned}
\right.
\end{equation}
which could make us feel confused owing to the difference with $|\Phi\rangle_{r}^{2}$ (The explanation  will be shown in Sec.\ref{analysis}).

\textbf{\emph{Case 2: position-dependent coins.}} Considering the fact that the coin unitary operator can be selected arbitrarily and it would generate various of  entangled states \cite{konno2013limit,zhang2014one,
suzuki2016asymptotic,ahmad2020one}. So whether it is possible by choosing the proper coin operator to obtain some special quantum state that is $|\Phi\rangle_{r}^{2}$ plus \emph{local} unitary operation. To do it, we have resorted to a position-dependent coin operator.

Taking $\mathbb{C}_{c_{1,2}}(1)=\mathbb{C}_{c_{1,3}}(2)=\mathbb{I}$, $\mathbb{C}_{c_{1,2}}(-1)=\mathbb{C}_{c_{1,3}}(-2)=\sigma_{x}$ as the coin operators ($\mathbb{C}_{c_{1,k}}(x)$ is the coin operation on the $k$-th coin whose action depends on the corresponding position $x$), thus after the second step $\mathcal{U}^{(2)}$, the state evolves into
\begin{equation}\label{flineN}
\begin{split}
\left|\Psi^{\prime\prime}_{p}\right\rangle&=\alpha\left|3\right\rangle\left|1\right\rangle^{\otimes (n-1)}\left|0\right\rangle^{\otimes n}\left|0\right\rangle\left|0\right\rangle+\beta\left|-3\right\rangle\left|-1\right\rangle^{\otimes (n-1)}\left|1\right\rangle^{\otimes n}\left|1\right\rangle\left|1\right\rangle,
\end{split}
\end{equation}
Compared with the basis on   $s_{p_{i}}$ in \textbf{\emph{Case 2}}, the measurement basis  turns to be $\Lambda_{p} = \{|\hspace{-1.5pt}-\hspace{-1.5pt}\hat{3}\rangle,|\hat{3}\rangle\}$,
with $|\hat{3}\rangle=(\left|3\right\rangle+\left|\hspace{-1pt}-\hspace{-1pt}3\right\rangle)/\sqrt{2},
|\hspace{-2pt}-\hspace{-2pt}\hat{3}\rangle=(\left|3\right\rangle-\left|\hspace{-1pt}-\hspace{-1pt}3\right\rangle)/\sqrt{2}$
, however, the measurement basis on  $s_{p_{k}}$ and $s_{c_{i,1}}$ denoted as $\Theta_{p}$ and $\Delta_{p}$ respectively remains unchanged, that is $\Theta_{p}=\Theta_{h}$ and $\Delta_{p}=\Delta_{h}$. Mark the measurements results 0 and 1 corresponding $|\hat{3}\rangle$ and $|\hspace{-2pt}-\hspace{-2pt}\hat{3}\rangle$ for $s_{p_{1}}$.  Therefore, the state owned by the receivers will be
\begin{equation}
|\Phi_{p}\rangle_{r}^{2}=\alpha|00\rangle+(-1)^{\omega_{p}}\beta|11\rangle.
\end{equation}
with  $\omega_{p}=\sum_{i=1}^{n}(p_{i}+c_{i})\, \text{mod}\, 2$. Then $\textbf{\emph{r}}_{1}$ or $\textbf{\emph{r}}_{2}$ can reconstruct  $|\Phi\rangle_{r}^{2}$ by the\emph{ local}  Pauli operator $\sigma_{z}^{\omega_{p}}$ at the location of $\textbf{\emph{r}}_{1}$ or $\textbf{\emph{r}}_{2}$.

\subsection{Scheme of teleportation between $n$ senders and $m$ receivers}
In Section \ref{n2}, the teleportation scheme between $n$ senders and $2$ receivers is explicitly demonstrated   by utilizing $n$-walker QWs, the first walker of which is driven by three coins. The next question is whether it is possible to implement the teleportation scheme between $n$ senders and $m$ receivers with QWs? Here, we generalize the  previous  $(n,2)$ scheme to the $(n,m)$ scheme
by increasing the amount of the coins of the first walker to $m+1$. The scheme is presented as follows.

The initial setting of the senders is analogy to the setting in the $(n,2)$ scheme. The difference takes place in the setting of the receivers. By adjusting  the number of the coins of the first walker to $m+1$, every receiver $\emph{\textbf{r}}_{j}(j=1,2,\cdots,m)$ separately has one particle $r_{c_{1,j+1}}$  corresponding to the $j\!+\!1$ coin of the
first walker. Thus, the joint Hilbert space of the composite system is $\mathcal{H}_{1,2,\cdots,n}=\otimes_ {i=1}^{n}\mathcal{H}_{p_i}
\otimes_{i=1}^{n}\mathcal{H}_{c_{i,1}}\otimes_{j=1}^{m}\mathcal{H}_{c_{1,j+1}}$ and we can derive the corresponding unitary operators of the two steps as follows

(i)The unitary operator of the first step is $\mathcal{U}^{(1)}_{m}=\mathcal{E}_{1}^{m}\mathcal{C}_{1}^{m}$ with
\begin{equation}\label{U1N}
\begin{split}
&\mathcal{C}_{1}^{m}=\otimes_{i=1}^{n}(\mathbb{I}_{p_{i}}\otimes\mathbb{C}_{c_{i,1}})
\otimes_{j=1}^{m}\mathbb{I}_{c_{1,j+1}},\\
&\mathcal{E}_{1}^{m}=\otimes_{i=1}^{n}(\mathcal{S}_{p_{i}}\otimes|0\rangle_{c_{i,1}}\langle 0|+\mathcal{S}_{p_{i}}^{\dag}\otimes|0\rangle_{c_{i,1}}\langle 0|)\otimes_{j=1}^{m}\mathbb{I}_{c_{1,j+1}},
\end{split}
\end{equation}

(ii)The unitary operator of the second step is $U^{(2)}=\Pi_{j=1}^{m} \mathcal{E}_{j+1}^{m}\mathcal{C}_{j+1}^{m}$ with
\begin{equation}\label{U2Nm}
\begin{split}
\mathcal{C}_{j+1}^{m}=&\otimes_{i=1}^{n}(\mathbb{I}_{p_{i}}\otimes\mathbb{I}_{c_{i,1}})
\otimes_{k=1}^{j-1}\mathbb{I}_{c_{1,k+1}}\otimes\mathbb{C}_{c_{1,j+1}}\otimes_{k=j+1}^{m}\mathbb{I}_{c_{1,k+1}},\\
\mathcal{E}_{j+1}^{m}=&\mathcal{S}_{p_{1}}\otimes\mathbb{I}_{c_{1,1}}\otimes_{i=2}^{n}(\mathbb{I}_{p_{i}}\otimes\mathbb{I}_{c_{i,1}})
\otimes_{k=1}^{j-1}\mathbb{I}_{c_{1,k+1}}|0\rangle_{c_{1,j+1}}\langle0|\otimes_{k=j+1}^{m}\mathbb{I}_{c_{1,k+1}}+\\
&\mathcal{S}_{p_{1}}^{\dag}\otimes\mathbb{I}_{c_{1,1}}\otimes_{i=2}^{n}(\mathbb{I}_{p_{i}}\otimes\mathbb{I}_{c_{i,1}})
\otimes_{k=1}^{j-1}\mathbb{I}_{c_{1,k+1}}|1\rangle_{c_{1,j+1}}\langle1|\otimes_{k=j+1}^{m}\mathbb{I}_{c_{1,k+1}}.
\end{split}
\end{equation}
Also,  the overall initial state can be written while assuming the coin state in $r_{c_{1,j+1}}$ is $\left|0\right\rangle$
\begin{equation}\label{initiaNNm}
\left|\Psi\right\rangle_{m}=\otimes_{i=1}^{n} \left|0\right\rangle_{p_{i}}\otimes(\alpha\otimes_{i=1}^{n}|0\rangle_{c_{i,1}}+\beta \otimes_{i=1}^{n}|1\rangle_{c_{i,1}})\otimes_{j=1}^{m}\left|0\right\rangle_{c_{1,j+1}},
\end{equation}
And in the first step, we keep taking $\mathbb{C}_{c_{i,1}}=\mathbb{I}$ as the coin operators, and the initial state (\ref{initiaNNm}) will become
\begin{equation}
\left|\Psi^{\prime}\right\rangle_{m}=\alpha\left|1\right\rangle^{\otimes n}\left|0\right\rangle^{\otimes n}\left|0\right\rangle^{\otimes m}
+\beta\left|-1\right\rangle^{\otimes n}\left|1\right\rangle^{\otimes n}\left|0\right\rangle^{\otimes m}.
\end{equation}
In order to accomplish the task, in analogy to the $(n,2)$ scheme, two kinds of coin operators that are the homogeneous coins and the position-dependent coins, are utilized to produce the entanglement state in the second step.

\textbf{\emph{Case 1: homogeneous coins.}}  We take $\mathbb{C}_{c_{1,j+1}}=\mathbb{H}$ as the coin operators, thus after the second step $U^{(2)}$, the state evolves into
\begin{small}
\begin{equation}\label{flineNNm}
\begin{split}
&\left|\Psi^{\prime\prime}_{h}\right\rangle_{m}=\sum_{k=0}^{m}(\alpha\left|m_{k}\right\rangle\left|1\right\rangle^{\otimes (n-1)}\left|0\right\rangle^{\otimes n}+\beta\left|m_{k+1}\right\rangle\left|-1\right\rangle^{\otimes (n-1)}\left|1\right\rangle^{\otimes n})\mathbf{P}[|1\rangle^{\otimes k}|0\rangle^{\otimes (m\!-\!k)}]\\
&=\underbrace{\sum_{k=0}^{m^{\prime\prime}}\alpha\left|m_{2k}\right\rangle\left|1\right\rangle^{\otimes (n-1)}\left|0\right\rangle^{\otimes n}\mathbf{P}[|1\rangle^{\otimes 2k}|0\rangle^{\otimes (m\!-\!2k\!-\!1)}]|0\rangle+\sum_{k=0}^{m^{\prime\prime}}\beta\left|m_{2k\!+\!2}\right\rangle\left|-1\right\rangle^{\otimes (n-1)}\left|1\right\rangle^{\otimes n}\mathbf{P}[|1\rangle^{\otimes (2k)}|0\rangle^{\otimes (m\!-\!2k-1)}]|1\rangle}\\
&+\underbrace{\sum_{k=1}^{m^{\prime}}\alpha\left|m_{2k}\right\rangle\left|1\right\rangle^{\otimes (n-1)}\left|0\right\rangle^{\otimes n}\mathbf{P}[|1\rangle^{\otimes (2k\!-\!1)}|0\rangle^{\otimes (m\!-\!2k)}]|1\rangle+\sum_{k=0}^{m^{\prime}-1}\beta\left|m_{2k\!+\!2}\right\rangle\left|-1\right\rangle^{\otimes (n-1)}\left|1\right\rangle^{\otimes n}\mathbf{P}[|1\rangle^{\otimes (2k+1)}|0\rangle^{\otimes (m\!-\!2k-2)}]|0\rangle}\\
&+\underbrace{\sum_{k=0}^{m^{\prime}\!-\!1}\alpha\left|m_{2k\!+\!1}\right\rangle\left|1\right\rangle^{\otimes (n-1)}\left|0\right\rangle^{\otimes n}\mathbf{P}[|1\rangle^{\otimes (2k\!+\!1)}|0\rangle^{\otimes (m\!-\!2k\!-\!2)}]|0\rangle+\sum_{k=1}^{m^{\prime}}\beta\left|m_{2k+1}\right\rangle\left|-1\right\rangle^{\otimes (n-1)}\left|1\right\rangle^{\otimes n}\mathbf{P}[|1\rangle^{\otimes (2k\!-\!1)}|0\rangle^{\otimes (m\!-\!2k)}]|1\rangle}\\
&+\underbrace{\sum_{k=0}^{m^{\prime\prime}}\alpha\left|m_{2k\!+\!1}\right\rangle\left|1\right\rangle^{\otimes (n-1)}\left|0\right\rangle^{\otimes n}\mathbf{P}[|1\rangle^{\otimes (2k)}|0\rangle^{\otimes (m\!-\!2k\!-\!1)}]|1\rangle+\sum_{k=0}^{m^{\prime\prime}}\beta\left|m_{2k+1}\right\rangle\left|-1\right\rangle^{\otimes (n-1)}\left|1\right\rangle^{\otimes n}\mathbf{P}[|1\rangle^{\otimes 2k}|0\rangle^{\otimes (m\!-\!2k\!-\!1)}]|0\rangle},
\end{split}
\end{equation}
\end{small}
where $m_{k}=m+1-2k$, $m^{\prime}=[\frac{m}{2}]$, $m^{\prime\prime}=[\frac{m-1}{2}]$, and $P[\cdot]$ denotes the sum of all possible permutation, for example, taking $m=3$ and $k=2$, $\mathbf{P}[|1\rangle|1\rangle|0\rangle]$ = $|011\rangle+|101\rangle+|110\rangle$ (See the appendix. A  for the calculation process of this formula (\ref{flineNNm})).  Then,  each sender $\textbf{\emph{s}}_{i}$ performs two single-particle measurements on $s_{p_{i}}$ and  $s_{c_{i,1}}$ with the corresponding basis
\begin{equation}\label{basis}
\left\{
\begin{aligned}
&\Lambda_{h}^{m}=\{|\tilde{\dot{m}}_{s}\rangle,|\tilde{\ddot{m}}_{t}\rangle|\,
\dot{m}_{s}=m\!+\!1\!-\!4s,\ddot{m}_{t}=m\!-\!1\!-\!4t,\,s,t \in Z, 0\leq \!s\!\leq m^{\prime\prime\prime}=m^{\prime\prime}\!+\!1,0\leq \!t\!\leq m^{\prime}\},\\
&\Theta_{h}^{m}=\Theta_{h};\,\,\Delta_{h}^{m}=\Delta_{h},
\end{aligned}
\right.
\end{equation}
with
\begin{subequations}\label{s1N}
\begin{align}
|\tilde{\dot{m}}_{s}\rangle&=\frac{1}{\sqrt{m^{\prime\prime\prime}+1}}\sum \nolimits_{l=0}^{m^{\prime\prime\prime}}e^{-2\pi i l s/(m^{\prime\prime\prime}+1)}|\dot{m}_{l}\rangle,\\
|\tilde{\ddot{m}}_{t}\rangle&=\frac{1}{\sqrt{m^{\prime}+1}}\sum \nolimits_{l=0}^{m^{\prime}}e^{-2\pi i l t/(m^{\prime}+1)}|\ddot{m}_{l}\rangle,
\end{align}
\end{subequations}
Mark the measurements results $0s$ and $1t$ corresponding  $|\tilde{\dot{m}}_{s}\rangle$ and $|\tilde{\ddot{m}}_{t}\rangle$ for $s_{p_{1}}$. Therefore, conditioned on the results, the final state obtained by the receivers can be written as
\begin{subnumcases}
{|\Phi_{h}\rangle_{r}^{m}=}\label{nms}
|\Phi_{h,s}\rangle_{r}^{m}: &$|1\rangle^{\otimes^{m\!-\!1}}_{\text{odd}}\sigma_{x}\sigma_{z}^{\omega_{h,m}}|\phi\rangle
+|1\rangle^{\otimes^{m\!-\!1}}_{\text{even}}\sigma_{z}^{\omega_{h,m}}R_{z}(\theta_{s})|\phi\rangle,\theta_{s}=2\pi s/(m^{\prime\prime\prime}+1),\,p_{1}=0s$,\\
|\Phi_{h,t}\rangle_{r}^{m}: &$|1\rangle^{\otimes^{m\!-\!1}}_{\text{even}}\sigma_{x}\sigma_{z}^{\omega_{h,m}}|\phi\rangle
+|1\rangle^{\otimes^{m\!-\!1}}_{\text{odd}}\sigma_{z}^{\omega_{h,m}}R_{z}(\vartheta_{t})|\phi\rangle,\vartheta_{t}=2\pi t/(m^{\prime}+1),\,p_{1}=1t$.
\end{subnumcases}
where $\omega_{h,m}=\omega_{h}$ and $R_{z}(\theta)$ is the rotation operator along  $z$ axis.

The analysis is similar with the $(n,2)$ scheme. In Lee {\it et} al.'s protocol\cite{lee2020quantum}, the final state at the receivers' locations is  $|\Phi\rangle_{r}^{m}$ plus \emph{logical} Pauli operations and the receivers can perform the  \emph{local} Pauli operations to retrieve $|\Phi\rangle_{r}^{m}$. The shared secret state $|\Phi\rangle_{s}^{m}$ can be rewritten as
\begin{equation}\label{shareds}
|\Phi\rangle_{s}^{m}=|-\rangle^{\otimes m\!-\!1}_{\text{even}}|\phi\rangle+|-\rangle_{\text{odd}}^{\otimes m\!-\!1}\sigma_{z}|\phi\rangle,
\end{equation}
with
\begin{equation}\label{sharedss}
\begin{split}
&|-\rangle^{\otimes m\!-\!1}_{\text{even(odd)}}=\sum_{j=\text{even(odd)}}^{m-1} \mathbf{P}[|-\rangle^{\otimes j}|+\rangle^{\otimes (m\!-\!1\!-\!j)}],
\end{split}
\end{equation}
e.g. $\mathbf{P}[|-\rangle|-\rangle|+\rangle]=|+--\rangle+|-+-\rangle+|--+\rangle$, while taking $m=4$ and $j=2$.

Next, taking the outcome $p_{1}=0s$ as an example to illustrate it. Then $\textbf{\emph{r}}_{m}$ performs $\sigma_{z}^{\omega_{h,m}}\sigma_{x}$ yielding
\begin{equation}\label{27}
|1\rangle^{\otimes^{m\!-\!1}}_{\text{odd}}|\phi\rangle
+(-1)^{\omega_{h,m}}|1\rangle^{\otimes^{m\!-\!1}}_{\text{even}}\sigma_{x}R_{z}(\theta_{s})|\phi\rangle,
\end{equation}
and then all of the other receivers $\{\textbf{\emph{r}}_{1},
\textbf{\emph{r}}_{2},\cdots,\textbf{\emph{r}}_{m-1}\}$ take the same Pauli operator  $\sigma_{z}^{\omega_{h,m}}$, which will make the expression (\ref{27})  become
\begin{equation}
|\Phi_{h,s}^{\prime}\rangle_{r}^{m}=|1\rangle^{\otimes^{m\!-\!1}}_{\text{odd}}|\phi\rangle
+|1\rangle^{\otimes^{m\!-\!1}}_{\text{even}}\sigma_{x}R_{z}(\theta_{s})|\phi\rangle.
\end{equation}
Note that $\textbf{\emph{r}}_{m}$ and all of the other receivers $\{\textbf{\emph{r}}_{1},
\textbf{\emph{r}}_{2},\cdots,\textbf{\emph{r}}_{m-1}\}$ could respectively also perform $R_{z}(-\theta_{s})\sigma_{z}^{\omega_{h,m}}$ and the Pauli operator  $\sigma_{z}^{\omega_{h,m}}$, which will arrive at $|1\rangle^{\otimes^{m\!-\!1}}_{\text{odd}}R_{z}(-\theta_{s})\sigma_{x}|\phi\rangle
+|1\rangle^{\otimes^{m\!-\!1}}_{\text{even}}|\phi\rangle$. However, the \emph{non-Pauli } operator $R_{z}(\theta_{s})$ is introduced, which might not be a good choice. Consequently, we take the former choice only with the Pauli operator.

Furthermore, considering the fact that
\begin{equation}
|a\rangle\sigma_{x}\sigma_{z}^{\omega_{h,m}}|\phi\rangle
+|(a\!+\!1)\,\text{mod}\, 2\rangle\sigma_{z}^{\omega_{h,m}}R_{z}(\theta_{s})|\phi\rangle=\sigma_{x}\sigma_{z}^{\omega_{h,m}}|\phi\rangle|a\rangle
+\sigma_{z}^{\omega_{h,m}}R_{z}(\theta_{s})|\phi\rangle|(a\!+\!1)\,\text{mod}\, 2\rangle,
\end{equation}
it can be concluded that any receivers $\textbf{\emph{r}}_{j}$ and the other $m-1$ receivers  can also obtain $|\Phi_{h,s}^{\prime}\rangle_{r}^{m}$ by performing the operators $\sigma_{z}^{\omega_{h,m}}\sigma_{x}$ and $\sigma_{z}^{\omega_{h,m}}$, respectively ((See the appendix. B for the explanation).  Similarly, while the result on $s_{p_{1}}$ is $0s$, the receivers can do the same  as   the result is $1t$, however, which will yield
\begin{equation}
|\Phi_{h,t}^{\prime}\rangle_{r}^{m}=|1\rangle^{\otimes^{m\!-\!1}}_{\text{even}}|\phi\rangle
+|1\rangle^{\otimes^{m\!-\!1}}_{\text{odd}}\sigma_{x}R_{z}(\vartheta_{t})|\phi\rangle,
\end{equation}

To sum up, in our $(n,m)$ scheme, after the QWs and the local Pauli operators, the initial shared quantum state $|\Phi\rangle_{s}^{n}$ become
\begin{equation}\label{phihrm}
|\Phi_{h}^{\prime}\rangle_{r}^{m}=\left\{
\begin{aligned}
&p_{1}=0s: |\Phi_{h,s}^{\prime}\rangle_{r}^{m},\\
&p_{1}=1t: |\Phi_{h,t}^{\prime}\rangle_{r}^{m}.
\end{aligned}
\right.
\end{equation}
which can also  be regarded as a new multiparty entanglement state to split the quantum secret state  $|\phi\rangle$ among $m$ parties, compared with the quantum state $|\Phi\rangle_{r}^{m}$. In addition,  the state $|\Phi^{\prime}_{h}\rangle_{r}^{m}$ is symmetric meaning that any receiver can reconstruct the quantum secret state with the help of the other receivers. Taking $m=2$, the $(n,m)$ scheme will degenerate into the previous  $(n,2)$ scheme in Sec.\ref{n2}.

\textbf{\emph{Case 2: position-dependent coins.}} Here, by choosing the proper \emph{position-dependent coin} operators, we can obtain some special quantum state that is local Pauli equivalent with $|\Phi\rangle_{s}^{m}$. Taking $\mathbb{C}_{c_{1,j+1}}(j)=\mathbb{I}$, $\mathbb{C}_{c_{1,j+1}}(-j)=\sigma_{x}$$(j=1,2,\cdots,m)$ as the coin operators, the state will be after the QWs
\begin{equation}\label{flineN2}
\begin{split}
\left|\Psi^{\prime\prime}_{p}\right\rangle_{m}&=\alpha\left|m\!+\!1\right\rangle\left|1\right\rangle^{\otimes (n-1)}\left|0\right\rangle^{\otimes n}\left|0\right\rangle^{\otimes m}+\beta\left|-(m\!+\!1)\right\rangle\left|-1\right\rangle^{\otimes (n-1)}\left|1\right\rangle^{\otimes n}\left|1\right\rangle^{\otimes m},
\end{split}
\end{equation}
The measurement basis  on $s_{p_{1}}$ will  be  $\Lambda_{p}^{m} = \{|\hspace{-1.5pt}-\hspace{-1.5pt}\hat{m}_{p}\rangle,|\hat{m}_{p}\rangle\}$
with $|\hat{m}_{p}\rangle=(\left|m_{p}\right\rangle+\left|\hspace{-1pt}-\hspace{-1pt}m_{p}\right\rangle)/\sqrt{2},
|\hspace{-2pt}-\hspace{-2pt}\hat{m}_{p}\rangle=(\left|m_{p}\right\rangle-\left|\hspace{-1pt}
-\hspace{-1pt}m_{p}\right\rangle)/\sqrt{2}$ and $m_{p}=m+1$,
and the measurement basis on the other $m-1$ walkers satisfy the condition $\Theta_{p}^{m}=\Theta_{h}^{m}$ and $\Delta_{p}^{m}=\Delta_{h}^{m}$. Mark the measurements results 0 and 1 corresponding $|\hat{m}_{p}\rangle$ and $|\hspace{-2pt}-\hspace{-2pt}\hat{m}_{p}\rangle$ for $s_{p_{1}}$.  Therefore, the state owned by the receivers will be
\begin{equation}
|\Phi_{p}\rangle_{r}^{m}=\alpha|0\rangle^{\otimes m}+(-1)^{\omega_{p,m}}\beta|1\rangle^{\otimes m}.
\end{equation}
with $\omega_{p,m}=\omega_{p}$. Then any receiver $\textbf{\emph{r}}_{j}$  can reconstruct  $|\Phi\rangle_{r}^{m}$ by the Pauli operator $\sigma_{z}^{\omega_{p,m}}$ at his location.

\section{Discussion}\label{discussion}
\subsection{Information security}\label{access}
The proposed protocol is aimed at achieving the goal that no single or subparties of the senders or receivers is allowed to access the secret during the whole process due to that none of the participants is fully trusted here.  The $n$ senders initially share a quantum secret $|\Phi\rangle_{s}^{n}=\alpha\otimes_{i=1}^{n}|0\rangle_{s_{i}}+\beta \otimes_{i=1}^{n}|1\rangle_{s_{i}}$. In our protocol, as soon as any subgroup of the senders, say $\{\textbf{\emph{s}}_{j_{1}},\textbf{\emph{s}}_{j_{2}},\cdots,\textbf{\emph{s}}_{j_{p}}\}(1 \leq p \leq n)$, perform the measurements, all the remaining $n\!+\!m\!-\!p$ participants, $\{\textbf{\emph{s}}_{j_{p+1}},\cdots,\textbf{\emph{s}}_{j_{n}},\textbf{\emph{r}}_{1},
\cdots,\textbf{\emph{r}}_{m}\}$
are entangled such that none of their sub-parties can fully access the shared secret. As presented above, we have proposed two kinds of $(n,m)$ teleportation protocols based on the different choices of the coin operator: \emph{position-dependent coins} and  \emph{homogeneous coins}, which are respectively denoted as \textbf{Protocol 1} and \textbf{Protocol 2} for brevity. Next, we will give the proof of the information security clearly and take $(n,2)$ scheme as an example to illustrate it in detail.

\textbf{Protocol 1}:
To begin with, we give the reduced  state of the remaining participants after the arbitrary $p$ senders $\{\textbf{\emph{s}}_{j_{1}},\textbf{\emph{s}}_{j_{2}},\cdots,\textbf{\emph{s}}_{j_{p}}\}$ carry out the measurements, which will bring out two kinds of results:

\textbf{case $1$}: $\textbf{\emph{s}}_{1} \notin \{\textbf{\emph{s}}_{j_{1}},\textbf{\emph{s}}_{j_{2}},\cdots,\textbf{\emph{s}}_{j_{p}}\}$. The corresponding reduced quantum state will be
\begin{small}
\begin{equation}\label{flineNNm1}
\begin{split}
\left|\Psi^{\prime\prime\prime}_{h_{1}}\right\rangle_{m}^{(n-p)}\!=\!\sum_{k=0}^{m}(\alpha\left|m_{k}\right\rangle\left|1\right\rangle^{\otimes (n-p-1)}\left|0\right\rangle^{\otimes (n-p)}\pm\beta\left|m_{k+1}\right\rangle\left|-1\right\rangle^{\otimes (n-p-1)}\left|1\right\rangle^{\otimes (n-p)})\mathbf{P}[|1\rangle^{\otimes k}|0\rangle^{\otimes (m-k)}],
\end{split}
\end{equation}
\end{small}By tracing out any sub-partites of $\{\textbf{\emph{s}}_{j_{p+1}}\!\!=\!\textbf{\emph{s}}_{1},\textbf{\emph{s}}_{j_{p+2}},\cdots,\textbf{\emph{s}}_{j_{n}},
\textbf{\emph{r}}_{1},\cdots,\textbf{\emph{r}}_{m}\}$, it will result in decoherence with the amplitude information only, so that the secret is not accessible to any single or sub-parties among the senders and receivers.

Next, we begin to minutely analyze the $(n,2)$ scheme. Taking $m=2$,  the reduced state of the remaining parties $\{\textbf{\emph{s}}_{1},\textbf{\emph{s}}_{j_{p+2}},\cdots,\textbf{\emph{s}}_{j_{n}},\textbf{\emph{r}}_{1},
\textbf{\emph{r}}_{2}\}$ is
\begin{equation}
\begin{split}
\left|\Psi^{\prime\prime\prime}_{h_{1}}\right\rangle_{2}^{n-p}&=\{\alpha\left|3\right\rangle\left|1\right\rangle^{\otimes (n-p-1)}\left|0\right\rangle^{\otimes (n-p)}\left|0\right\rangle\left|0\right\rangle
+\alpha\left|1\right\rangle\left|1\right\rangle^{\otimes (n-p-1)}\left|0\right\rangle^{\otimes (n-p)}\left|0\right\rangle\left|1\right\rangle+\\
&\alpha\left|1\right\rangle\left|1\right\rangle^{\otimes (n-p-1)}\left|0\right\rangle^{\otimes (n-p)}\left|1\right\rangle\left|0\right\rangle
+\alpha\left|-1\right\rangle\left|1\right\rangle^{\otimes (n-p-1)}\left|0\right\rangle^{\otimes (n-p)}\left|1\right\rangle\left|1\right\rangle\}\pm\\
&\{\beta\left|1\right\rangle\left|-1\right\rangle^{\otimes (n-p-1)}\left|1\right\rangle^{\otimes (n-p)}\left|0\right\rangle\left|0\right\rangle+\beta\left|-1\right\rangle\left|-1\right\rangle^{\otimes (n-p-1)}\left|1\right\rangle^{\otimes (n-p)}\left|0\right\rangle\left|1\right\rangle+\\
&\beta\left|-1\right\rangle\left|-1\right\rangle^{\otimes (n-p-1)}\left|1\right\rangle^{\otimes (n-p)}\left|1\right\rangle\left|0\right\rangle+\beta\left|-3\right\rangle\left|-1\right\rangle^{\otimes (n-p-1)}\left|1\right\rangle^{\otimes (n-p)}\left|1\right\rangle\left|1\right\rangle\},
\end{split}
\end{equation}
which can be derived that only the amplitude information is accessible by tracing out any sub-parties. For example,  tracing out the sub-group $\{\textbf{\emph{s}}_{1},\textbf{\emph{s}}_{j_{p+2}},\cdots,\textbf{\emph{s}}_{j_{n}}\}$ will yield a mixed quantum state  $\mathbb{I}_{4}+|01\rangle\langle10|+|10\rangle\langle01|$, with the amplitude information only. Similarly, tracing out any sub-parties will yield the same result.

\textbf{case $2$}: $\textbf{\emph{s}}_{1} \in \{\textbf{\emph{s}}_{j_{1}},\textbf{\emph{s}}_{j_{2}},\cdots,\textbf{\emph{s}}_{j_{p}}\}$. As described by the expression (\ref{s1N}), two kinds of measurement basis sets are taken which will result in two corresponding quantum states below
\begin{subnumcases}
{\left|\Psi^{\prime\prime\prime}_{h_{2}}\right\rangle_{m}^{n-p}=}\notag
|\tilde{\dot{m}}_{s}\rangle: \sum_{k=0}^{m^{\prime}}\alpha\, e^{2\pi i k s/(m^{\prime\prime\prime}+1)}\left|1\right\rangle^{\otimes (n-p)}\left|0\right\rangle^{\otimes (n-p)}\mathbf{P}[|1\rangle^{\otimes 2k}|0\rangle^{\otimes (m-2k)}]\\
\pm\sum_{k=0}^{m^{\prime\prime}}\beta\, e^{2\pi i (k\!+\!1)  s /(m^{\prime\prime\prime}+1)}\left|-1\right\rangle^{\otimes (n-p)}\left|1\right\rangle^{\otimes (n-p)}\mathbf{P}[|1\rangle^{\otimes (2k+1)}|0\rangle^{\otimes (m-2k-1)}],\\\notag
|\tilde{\ddot{m}}_{t}\rangle: \sum_{k=0}^{m^{\prime\prime}}\alpha e^{2\pi i k t/(m^{\prime}+1)}\left|1\right\rangle^{\otimes (n-p)}\left|0\right\rangle^{\otimes (n-p)}\mathbf{P}[|1\rangle^{\otimes 2k+1}|0\rangle^{\otimes(m\!-\!2k\!-\!1)}]\\
\pm\sum_{k=1}^{m^{\prime}}\alpha e^{2\pi i (k+1) t/(m^{\prime}+1)}\left|1\right\rangle^{\otimes (n-p)}\left|0\right\rangle^{\otimes (n-p)}\mathbf{P}[|1\rangle^{\otimes 2k}|0\rangle^{\otimes (m\!-\!2k)}]
\end{subnumcases}
It can also be proved that  only the amplitude information is accessible while tracing out any sub-partites of $\{\textbf{\emph{s}}_{j_{p+1}},\textbf{\emph{s}}_{j_{p+2}},\cdots,\textbf{\emph{s}}_{j_{n}},
\textbf{\emph{r}}_{1},\cdots,\textbf{\emph{r}}_{m}\}$, which means that the secret is not accessible to any single or sub-parties among the senders and receivers.

Next, we take $(n,2)$ scheme to explain it in detail. Here, without loss of generality,  we take the case $|\tilde{\dot{m}}_{s}\rangle$ as the measurement basis to show it.  In this case, $m^{\prime}=1$, $m^{\prime\prime}=0$, $m^{\prime\prime\prime}=1$,  $s=\{0,1\}$, therefore, the reduced state at the remain parties will become
\begin{equation}
\begin{split}
\left|\Psi^{\prime\prime\prime}_{h_{2}}\right\rangle_{2}^{n-p}&=\{\alpha\left|1\right\rangle^{\otimes (n-p)}\left|0\right\rangle^{\otimes (n-p)}\left|0\right\rangle\left|0\right\rangle
+\alpha(-1)^{s}\left|1\right\rangle^{\otimes (n-p)}\left|0\right\rangle^{\otimes (n-p)}\left|1\right\rangle\left|1\right\rangle\}\pm\\
&\{\beta(-1)^{s}\left|-1\right\rangle^{\otimes (n-p)}\left|1\right\rangle^{\otimes (n-p)}\left|0\right\rangle\left|1\right\rangle+\beta(-1)^{s}\left|-1\right\rangle^{\otimes (n-p)}\left|1\right\rangle^{\otimes (n-p)}\left|1\right\rangle\left|0\right\rangle\},
\end{split}
\end{equation}
for example,  by tracing out the sub-group $\{\textbf{\emph{s}}_{j_{p+1}},\textbf{\emph{s}}_{j_{p+2}},\cdots,\textbf{\emph{s}}_{j_{n}}\}$, it will yield a mixed quantum state
\begin{small}
\begin{equation}|\alpha|^{2}[|00\rangle\langle00|+|11\rangle\langle11|+(-1)^{s}(|00\rangle\langle11|
+|11\rangle\langle00|)]+|\beta|^{2}[|01\rangle\langle01|+|10\rangle\langle10|+(-1)^{s}(|01\rangle\langle10|
+|10\rangle\langle01|)].\end{equation}\end{small}with the amplitude information only. Similarly, the same holds for any subparties of senders and receivers for information security.

\textbf{Protocol 2}:  The accessible information in \textbf{Protocol 1} has been analyzed in detail. Here, we will show the result  in \textbf{Protocol 2} directly. Firstly, according to the expression (\ref{flineN2}), we give the following reduced quantum state of all the remaining $n\!+\!m\!-\!p$ participants $\{\textbf{\emph{s}}_{j_{p+1}},\cdots,\textbf{\emph{s}}_{j_{n}},\textbf{\emph{r}}_{1},
\cdots,\textbf{\emph{r}}_{m}\}$
\begin{subnumcases}
{}
\left|\Psi^{\prime\prime\prime}_{p_{1}}\right\rangle_{m}^{n-p}=\alpha\left|m\!+\!1\right\rangle\left|1\right\rangle^{\otimes (n-p-1)}\left|0\right\rangle^{\otimes (n-p)}\left|0\right\rangle^{\otimes m}\pm\beta\left|-(m\!+\!1)\right\rangle\left|-1\right\rangle^{\otimes (n-p-1)}\left|1\right\rangle^{\otimes (n-p)}\left|1\right\rangle^{\otimes m},\\
\left|\Psi^{\prime\prime\prime}_{p_{2}}\right\rangle_{m}^{n-p}=\alpha\left|1\right\rangle^{\otimes (n-p)}\left|0\right\rangle^{\otimes (n-p)}\left|0\right\rangle^{\otimes m}\pm\beta\left|-1\right\rangle^{\otimes (n-p)}\left|1\right\rangle^{\otimes (n-p)}\left|1\right\rangle^{\otimes m},
\end{subnumcases}
where $\left|\Psi^{\prime\prime\prime}_{p_{1}}\right\rangle_{m}^{n-p}$ and $\left|\Psi^{\prime\prime\prime}_{p_{2}}\right\rangle_{m}^{n-p}$ are similar with $\left|\Psi^{\prime\prime\prime}_{h_{1}}\right\rangle_{m}^{n-p}$ and $\left|\Psi^{\prime\prime\prime}_{h_{2}}\right\rangle_{m}^{n-p}$, respectively. Specially, taking $m=2$, the corresponding reduced quantum state
\begin{subnumcases}
{}
\left|\Psi^{\prime\prime\prime}_{p_{1}}\right\rangle_{2}^{n-p}=\alpha\left|3\right\rangle\left|1\right\rangle^{\otimes (n-p-1)}\left|0\right\rangle^{\otimes (n-p)}\left|0\right\rangle^{\otimes 2}\pm\beta\left|-3\right\rangle\left|-1\right\rangle^{\otimes (n-p-1)}\left|1\right\rangle^{\otimes (n-p)}\left|1\right\rangle^{\otimes 2},\\
\left|\Psi^{\prime\prime\prime}_{p_{2}}\right\rangle_{2}^{n-p}=\alpha\left|1\right\rangle^{\otimes (n-p)}\left|0\right\rangle^{\otimes (n-p)}\left|0\right\rangle^{\otimes 2}\pm\beta\left|-1\right\rangle^{\otimes (n-p)}\left|1\right\rangle^{\otimes (n-p)}\left|1\right\rangle^{\otimes 2}.
\end{subnumcases}
By tracing out any sub-parties of $\{\textbf{\emph{s}}_{p+1},\textbf{\emph{s}}_{j_{p+2}},\cdots,\textbf{\emph{s}}_{j_{n}},
\textbf{\emph{r}}_{1},\cdots,\textbf{\emph{r}}_{m}\}$, it will also result in decoherence with the amplitude information only, so that the secret is not accessible to  any sub-parties of senders and receivers.

Above all,  we have proved that any subparties cannot fully
access the quantum secret during the teleportation procedures in the proposed $(n,m)$ schemes \textbf{Protocol 1} and \textbf{Protocol 2}.

\subsection{The analysis of our scheme} \label{analysis}
Multipartite entangled states are a fundamental resource for a wide range of quantum information processing tasks and we accomplished the transfer of the multipartite entangled states from $n$ senders to $m$ receivers. The $(n,2)$ teleportation scheme was firstly clearly presented via the model of  $n$-walker QWs, the first walker of which is driven by three coins and then it can be   generalized  to the $(n,m)$ scheme by increasing the amount of the coins of the first walker to $m+1$. Specially, $n$ senders $\textbf{\emph{s}}_{i}(1 \leq i \leq n)$  hold $2n$ particles, the corresponding position  and the first coin state of $n$ walkers, and  the shared secret quantum state $|\Phi\rangle_{s}^{n}$ is encoded in the coin states. Similarly, each receiver $\textbf{\emph{r}}_{i}(1 \leq i \leq m)$ has one particle $r_{c_{1,i+1}}$ corresponding to the $i\!+\!1$ coin of the first walker. We note that our work is not limited to QWs on the line and it can be extended further to QWs on the cycle \cite{aharonov2001quantum}.

Neither any single nor subparties of the participants can fully access the secret quantum information  in our proposed scheme, which has been proved in Sec.\ref{access}. It  implies that the participants  can relay quantum information over a network without requiring fully trusted central or intermediate nodes, which might be useful to establish a long-distance quantum communication via distributed nodes, none of which necessarily relays the full quantum information. Compared with Lee {\it et} al.'s protocol\cite{lee2020quantum},  single-particle measurements are needed, instead of joint measurements. In particular,  $2n$ times of projective  measurements are needed in our scheme and the measurement basis are $\{\Lambda_{h}^{m},\Theta_{h}^{m},\Delta_{h}^{m}\}$ and $\{\Lambda_{p}^{m},\Theta_{p}^{m},\Delta_{p}^{m}\}$, respectively corresponding \textbf{Protocol 1} and \textbf{Protocol 2}. However, $n$ times of standard Bell-state measurements are needed, in which a distributed Bell-state measurement is introduced that can be jointly performed by separated parties \cite{lee2020quantum}.

In \textbf{Protocol 1}, the initial shared quantum state $|\Phi\rangle_{s}^{n}$ become $|\Phi^{\prime}_{h}\rangle_{r}^{m}$ instead of $|\Phi\rangle_{r}^{m}$, which might seem that the teleportation task of the quantum state $|\Phi\rangle_{s}^{n}$ failed. Now let us turn our attention to the original motivation for the scheme, which is aimed at transferring the shared quantum  secret information $|\phi\rangle=\alpha|0\rangle_{L}+\beta|1\rangle_{L}$  shared by $n$ senders through a splitting protocol. It is obviously that the state  $|\Phi^{\prime}_{h}\rangle_{r}^{m}$ meets the requirement according to the formula (\ref{phihrm}), which can be regarded as a new entanglement channel to split the quantum secret state  $|\phi\rangle$. Therefore, the change of the  teleported shared quantum state  is not irrelevant for the  transmission of the shared quantum secret in \textbf{Protocol 1}. However, it would not take place while introducing \emph{position-dependent coins} in \textbf{Protocol 2}. By choosing the proper coin operators depending on the position, the receivers can obtain the reduced state $|\Phi_{p}\rangle_{r}^{m}=|\Phi\rangle_{r}^{m}$ plus logical Pauli operators, meaning that the receivers can carry out the local Pauli operators to retrieve  $|\Phi\rangle_{r}^{m}$.

During the \emph{recovery} process, conditioned on all the measurement results, the receivers  can carry out the local Pauli operators to retrieve the initial state quantum state, specially (i) in \textbf{Protocol 1}, any one of the receivers $\textbf{\emph{r}}_{j}$ needs to perform the unitary operator $\sigma_{z}^{\omega_{h,m}}\sigma_{x}$, and all the other ones need to perform the unitary operator $\sigma_{z}^{\omega_{h,m}}$; (ii) in \textbf{Protocol 2}, only any one of the receivers need to take the Pauli operator $\sigma_{z}^{\omega_{p,m}}$ at his location. In addition,  both  $|\Phi^{\prime}_{h}\rangle_{r}^{m}$ and $|\Phi^{\prime}_{p}\rangle_{r}^{m}$ are symmetric meaning that any receiver can reconstruct the quantum secret state with the help of the other receivers.  Moreover, as the logical outcome is irrespective of the order of performed measurements,  it is possible to separate all the measurements spatially or temporally with the help of classical communications to share their results among the nodes.

Additionally, as shown in Lee {\it et} al.'s protocol\cite{lee2020quantum}, the GHZ-entanglement  state $|\Phi\rangle^{(n+m)}=|0\rangle^{\otimes (n+m)}+|1\rangle^{\otimes (n+m)}$ is utilized as the network channel to fulfill the teleportation scheme. Due to  QWs can contain entanglement by the conditional shift operator between the walkers and the coins \cite{Brun2003Quantum,Omar2006Quantum},  the entangled state is not necessarily prepared in advance in our schemes. Furthermore, conditioned on the various choices of
the coin unitary operator, different kinds of entanglement sources can be generated to meet our requirements. In \textbf{Protocol 1},  the model of  \emph{the homogeneous coins} is used and we take $\mathbb{H}$ as the coin operator for producing the maximal entanglement state, which is the unbiased coin flipping operator and can  change $|0\rangle $ to $|0\rangle+|1\rangle$ with the same amplitude, and therefore meets the requirement.  In \textbf{Protocol 2}, for obtaining the specific quantum state, we introduce \emph{position-dependent coins} and show an approach to solving it, that is $\mathbb{C}_{c_{1,j+1}}(j)=\mathbb{I}$, $\mathbb{C}_{c_{1,j+1}}(-j)=\sigma_{x}$$(j=1,2,\cdots,m)$.

\section{Conclusion}
\label{Conclusion}
In this manuscript, by using the model of multi-walker QWs with multiple coins and two different coin operators, the \emph{homogeneous coin operators} and the \emph{position-dependent coin operators}, two kinds of corresponding novel $(n,m)$ teleportation schemes of shared quantum secret state among  $n$ senders and $m$ receivers, were proposed. This work thus not only opens a route to the realization of secure distributed quantum communications and computations in quantum networks, but also  provides an additional relevant instance of the richness of QWs for quantum information processing tasks and thus opens the wider application purpose  of quantum walks.

\section*{Acknowledgements}
This work was supported  by the National Natural Science Foundation of China (Grant Nos.U1636106, 61671087, 61170272), the BUPT Excellent Ph.D. Students Foundation(No.CX2020310
), Natural Science Foundation of Beijing Municipality (No.4182006), Technological Special Project of Guizhou Province (Grant No. 20183001), and the Foundation of Guizhou Provincial Key Laboratory of Public Big Data (Grant No.2018BDKFJJ016, 2018BDKFJJ018) and  the Fundamental Research Funds for the Central Universities (No.2019XD-A02).

\section*{Appendix}

\subsection*{Appendix A. The derivation process of the formula (\ref{flineNNm})}\label{app1}
\setcounter{equation}{0}
\setcounter{subsection}{0}
\renewcommand{\theequation}{A.\arabic{equation}}
\renewcommand{\thesubsection}{A.\arabic{subsection}}
Firstly, according the parity of $k$,  the  quantum state $\left|\Phi^{\prime\prime}\right\rangle$ can be written as $\left|\Phi^{\prime\prime}\right\rangle=
\left|\Phi^{\prime\prime}\right\rangle_{\text{even}}
+\left|\Phi^{\prime\prime}\right\rangle_{\text{odd}}$, defined by
\begin{subequations}
\begin{align}
&\left|\Phi^{\prime\prime}\right\rangle_{\text{even}}=\sum_{k=0}^{m^{\prime}}\alpha\left|m_{2k}\right\rangle\left|1\right\rangle^{\otimes (n-1)}\left|0\right\rangle^{\otimes n}\mathbf{P}[|1\rangle^{\otimes 2k}|0\rangle^{\otimes (m\!-\!2k)}]+\sum_{k=0}^{m^{\prime\prime}}\beta\left|m_{2k\!+\!2}\right\rangle\left|-1\right\rangle^{\otimes (n-1)}\left|1\right\rangle^{\otimes n}\mathbf{P}[|1\rangle^{\otimes (2k\!+\!1)}|0\rangle^{\otimes (m\!-\!2k\!-\!1)}]\\
&\left|\Phi^{\prime\prime}\right\rangle_{\text{odd}}=\sum_{k=0}^{m^{\prime\prime}}\alpha\left|m_{2k\!+\!1}\right\rangle\left|1\right\rangle^{\otimes (n-1)}\left|0\right\rangle^{\otimes n}\mathbf{P}[|1\rangle^{\otimes (2k\!+\!1)}|0\rangle^{\otimes (m\!-\!2k\!-\!1)}]+\sum_{k=0}^{m^{\prime}}\beta\left|m_{2k+1}\right\rangle\left|-1\right\rangle^{\otimes (n-1)}\left|1\right\rangle^{\otimes n}\mathbf{P}[|1\rangle^{\otimes 2k}|0\rangle^{\otimes (m\!-\!2k)}]
\end{align}
\end{subequations}
where $m^{\prime}=[\frac{m}{2}]$, $m^{\prime\prime}=[\frac{m-1}{2}]$. Next, by utilizing the following equations:
\begin{subequations}
\begin{align}
&\mathbf{P}[|1\rangle^{\otimes 2k}|0\rangle^{\otimes (m\!-\!2k)}]=\mathbf{P}[|1\rangle^{\otimes 2k}|0\rangle^{\otimes (m\!-\!2k\!-\!1)}]|0\rangle+
\mathbf{P}[|1\rangle^{\otimes 2k\!-\!1}|0\rangle^{\otimes (m\!-\!2k)}]|1\rangle\\
&\mathbf{P}[|1\rangle^{\otimes 2k\!+\!1}|0\rangle^{\otimes (m\!-\!2k\!-\!1)}]=\mathbf{P}[|1\rangle^{\otimes 2k\!+\!1}|0\rangle^{\otimes (m\!-\!2k\!-\!2)}]|0\rangle+
\mathbf{P}[|1\rangle^{\otimes 2k}|0\rangle^{\otimes (m\!-\!2k\!-\!1)}]|1\rangle
\end{align}
\end{subequations}
the state $\left|\Phi^{\prime\prime}\right\rangle$ can be rewritten as
\begin{equation}
\begin{split}
&\left|\Phi^{\prime\prime}\right\rangle=\sum_{k=0}^{m^{\prime\prime}}\alpha\left|m_{2k}\right\rangle\left|1\right\rangle^{\otimes (n-1)}\left|0\right\rangle^{\otimes n}\mathbf{P}[|1\rangle^{\otimes 2k}|0\rangle^{\otimes (m\!-\!2k\!-\!1)}]|0\rangle+\sum_{k=1}^{m^{\prime}}\alpha\left|m_{2k}\right\rangle\left|1\right\rangle^{\otimes (n-1)}\left|0\right\rangle^{\otimes n}\mathbf{P}[|1\rangle^{\otimes (2k\!-\!1)}|0\rangle^{\otimes (m\!-\!2k)}]|1\rangle\\
&+\sum_{k=0}^{m^{\prime}-1}\beta\left|m_{2k\!+\!2}\right\rangle\left|-1\right\rangle^{\otimes (n-1)}\left|1\right\rangle^{\otimes n}\mathbf{P}[|1\rangle^{\otimes (2k+1)}|0\rangle^{\otimes (m\!-\!2k-2)}]|0\rangle+\sum_{k=0}^{m^{\prime\prime}}\beta\left|m_{2k\!+\!2}\right\rangle\left|-1\right\rangle^{\otimes (n-1)}\left|1\right\rangle^{\otimes n}\mathbf{P}[|1\rangle^{\otimes (2k)}|0\rangle^{\otimes (m\!-\!2k-1)}]|1\rangle\\
&+\sum_{k=0}^{m^{\prime}\!-\!1}\alpha\left|m_{2k\!+\!1}\right\rangle\left|1\right\rangle^{\otimes (n-1)}\left|0\right\rangle^{\otimes n}\mathbf{P}[|1\rangle^{\otimes (2k\!+\!1)}|0\rangle^{\otimes (m\!-\!2k\!-\!2)}]|0\rangle+\sum_{k=0}^{m^{\prime\prime}}\alpha\left|m_{2k\!+\!1}\right\rangle\left|1\right\rangle^{\otimes (n-1)}\left|0\right\rangle^{\otimes n}\mathbf{P}[|1\rangle^{\otimes (2k)}|0\rangle^{\otimes (m\!-\!2k\!-\!1)}]|1\rangle\\
&+\sum_{k=0}^{m^{\prime\prime}}\beta\left|m_{2k+1}\right\rangle\left|-1\right\rangle^{\otimes (n-1)}\left|1\right\rangle^{\otimes n}\mathbf{P}[|1\rangle^{\otimes 2k}|0\rangle^{\otimes (m\!-\!2k\!-\!1)}]|0\rangle+\sum_{k=1}^{m^{\prime}}\beta\left|m_{2k+1}\right\rangle\left|-1\right\rangle^{\otimes (n-1)}\left|1\right\rangle^{\otimes n}\mathbf{P}[|1\rangle^{\otimes (2k\!-\!1)}|0\rangle^{\otimes (m\!-\!2k)}]|1\rangle\\
\end{split}
\end{equation}
Then, by adjusting the order of the terms, it will become
\begin{equation}
\begin{split}
&\left|\Phi^{\prime\prime}\right\rangle=\underbrace{\sum_{k=0}^{m^{\prime\prime}}\alpha\left|m_{2k}\right\rangle\left|1\right\rangle^{\otimes (n-1)}\left|0\right\rangle^{\otimes n}\mathbf{P}[|1\rangle^{\otimes 2k}|0\rangle^{\otimes (m\!-\!2k\!-\!1)}]|0\rangle+\sum_{k=0}^{m^{\prime\prime}}\beta\left|m_{2k\!+\!2}\right\rangle\left|-1\right\rangle^{\otimes (n-1)}\left|1\right\rangle^{\otimes n}\mathbf{P}[|1\rangle^{\otimes (2k)}|0\rangle^{\otimes (m\!-\!2k-1)}]|1\rangle}\\
&+\underbrace{\sum_{k=1}^{m^{\prime}}\alpha\left|m_{2k}\right\rangle\left|1\right\rangle^{\otimes (n-1)}\left|0\right\rangle^{\otimes n}\mathbf{P}[|1\rangle^{\otimes (2k\!-\!1)}|0\rangle^{\otimes (m\!-\!2k)}]|1\rangle+\sum_{k=0}^{m^{\prime}-1}\beta\left|m_{2k\!+\!2}\right\rangle\left|-1\right\rangle^{\otimes (n-1)}\left|1\right\rangle^{\otimes n}\mathbf{P}[|1\rangle^{\otimes (2k+1)}|0\rangle^{\otimes (m\!-\!2k-2)}]|0\rangle}\\
&+\underbrace{\sum_{k=0}^{m^{\prime}\!-\!1}\alpha\left|m_{2k\!+\!1}\right\rangle\left|1\right\rangle^{\otimes (n-1)}\left|0\right\rangle^{\otimes n}\mathbf{P}[|1\rangle^{\otimes (2k\!+\!1)}|0\rangle^{\otimes (m\!-\!2k\!-\!2)}]|0\rangle+\sum_{k=1}^{m^{\prime}}\beta\left|m_{2k+1}\right\rangle\left|-1\right\rangle^{\otimes (n-1)}\left|1\right\rangle^{\otimes n}\mathbf{P}[|1\rangle^{\otimes (2k\!-\!1)}|0\rangle^{\otimes (m\!-\!2k)}]|1\rangle}\\
&+\underbrace{\sum_{k=0}^{m^{\prime\prime}}\alpha\left|m_{2k\!+\!1}\right\rangle\left|1\right\rangle^{\otimes (n-1)}\left|0\right\rangle^{\otimes n}\mathbf{P}[|1\rangle^{\otimes (2k)}|0\rangle^{\otimes (m\!-\!2k\!-\!1)}]|1\rangle+\sum_{k=0}^{m^{\prime\prime}}\beta\left|m_{2k+1}\right\rangle\left|-1\right\rangle^{\otimes (n-1)}\left|1\right\rangle^{\otimes n}\mathbf{P}[|1\rangle^{\otimes 2k}|0\rangle^{\otimes (m\!-\!2k\!-\!1)}]|0\rangle}
\end{split}
\end{equation}
which is the same with the formula (\ref{flineNNm}).

\subsection*{Appendix B.}\label{app2}
\setcounter{equation}{0}
\setcounter{subsection}{0}
\renewcommand{\theequation}{B.\arabic{equation}}
\renewcommand{\thesubsection}{B.\arabic{subsection}}

To begin with, we give the following equation
\begin{equation}\label{appB1}
\begin{split}
|0\rangle\sigma_{x}\sigma_{z}^{\omega_{h,m}}|\phi\rangle
+|1\rangle\sigma_{z}^{\omega_{h,m}}R_{z}(\theta_{s})|\phi\rangle
&=|0\rangle[\alpha|1\rangle+(-1)^{\omega_{h,m}}\beta|0\rangle]
+|1\rangle[\alpha|0\rangle+(-1)^{\omega_{h,m}} e^{i \theta_{s}}\beta|1\rangle]\\
&=[\alpha|1\rangle+(-1)^{\omega_{h,m}}\beta|0\rangle]|0\rangle
+[\alpha|0\rangle+(-1)^{\omega_{h,m}} e^{i \theta_{s}}\beta|1\rangle]|1\rangle\\
&=\sigma_{x}\sigma_{z}^{\omega_{h,m}}|\phi\rangle|0\rangle
+\sigma_{z}^{\omega_{h,m}}R_{z}(\theta_{s})|\phi\rangle|1\rangle,
\end{split}
\end{equation}
Similarly, it can be easily deduced  that
\begin{equation}\label{appB2}
\begin{split}
|1\rangle\sigma_{x}\sigma_{z}^{\omega_{h,m}}|\phi\rangle
+|0\rangle\sigma_{z}^{\omega_{h,m}}R_{z}(\theta_{s})|\phi\rangle
=\sigma_{x}\sigma_{z}^{\omega_{h,m}}|\phi\rangle|1\rangle
+\sigma_{z}^{\omega_{h,m}}R_{z}(\theta_{s})|\phi\rangle|0\rangle,
\end{split}
\end{equation}

Next, we will prove that the quantum states of $\textbf{\emph{r}}_{m-1}$ and $\textbf{\emph{r}}_{m}$ are \emph{exchangeable}, that is, the composite state keeps constant after carrying out the Swap gate between $\textbf{\emph{r}}_{m-1}$ and $\textbf{\emph{r}}_{m}$.  Recall $|\Phi_{h,s}\rangle_{r}^{m}$ in (\ref{nms}) and by using the (\ref{appB1}) and (\ref{appB2}), it  can be rewritten as
\begin{equation}
\begin{split}
|\Phi_{h,s}\rangle_{r}^{m}&=|1\rangle^{\otimes^{m\!-\!1}}_{\text{odd}}\sigma_{x}\sigma_{z}^{\omega_{h,m}}|\phi\rangle
+|1\rangle^{\otimes^{m\!-\!1}}_{\text{even}}\sigma_{z}^{\omega_{h,m}}R_{z}(\theta_{s})|\phi\rangle\\
&=|1\rangle^{\otimes^{m\!-\!2}}_{\text{odd}}|0\rangle\sigma_{x}\sigma_{z}^{\omega_{h,m}}|\phi\rangle
+|1\rangle^{\otimes^{m\!-\!2}}_{\text{even}}|1\rangle\sigma_{x}\sigma_{z}^{\omega_{h,m}}|\phi\rangle+|1\rangle^{\otimes^{m\!-\!2}}_{\text{even}}|0\rangle\sigma_{z}^{\omega_{h,m}}R_{z}(\theta_{s})|\phi\rangle
+|1\rangle^{\otimes^{m\!-\!2}}_{\text{odd}}|1\rangle\sigma_{z}^{\omega_{h,m}}R_{z}(\theta_{s})|\phi\rangle\\
&=|1\rangle^{\otimes^{m\!-\!2}}_{\text{odd}}(|0\rangle\sigma_{x}\sigma_{z}^{\omega_{h,m}}|\phi\rangle
+|1\rangle\sigma_{z}^{\omega_{h,m}}R_{z}(\theta_{s})|\phi\rangle)+|1\rangle^{\otimes^{m\!-\!2}}_{\text{even}}(|1\rangle\sigma_{x}\sigma_{z}^{\omega_{h,m}}|\phi\rangle
+|0\rangle\sigma_{z}^{\omega_{h,m}}R_{z}(\theta_{s})|\phi\rangle)\\
&=|1\rangle^{\otimes^{m\!-\!2}}_{\text{odd}}(\sigma_{x}\sigma_{z}^{\omega_{h,m}}|\phi\rangle|0\rangle
+\sigma_{z}^{\omega_{h,m}}R_{z}(\theta_{s})|\phi\rangle)|1\rangle+|1\rangle^{\otimes^{m\!-\!2}}_{\text{even}}(\sigma_{x}\sigma_{z}^{\omega_{h,m}}|\phi\rangle|1\rangle
+\sigma_{z}^{\omega_{h,m}}R_{z}(\theta_{s})|\phi\rangle|0\rangle)
\end{split}
\end{equation}
which says that the quantum states of $\textbf{\emph{r}}_{m-1}$ and $\textbf{\emph{r}}_{m}$ are \emph{exchangeable}. It can be  obviously concluded that any two receivers of
$\{\textbf{\emph{r}}_{1},\cdots,\textbf{\emph{r}}_{m-1}\}$ are also \emph{exchangeable} owing to the symmetry of   $|1\rangle^{\otimes^{m\!-\!1}}_{\text{odd}}$ and $|1\rangle^{\otimes^{m\!-\!1}}_{\text{even}}$. Thus, the  quantum states of $\textbf{\emph{r}}_{j}$ and $\textbf{\emph{r}}_{m}$ are \emph{exchangeable} and then it will give
\begin{equation}
\begin{split}
|\Phi_{h,s}\rangle_{r}^{m}=
|1\rangle^{\otimes^{j\!-\!1}}_{\text{even}}\sigma_{x}\sigma_{z}^{\omega_{h,m}}|\phi\rangle_{j}|1\rangle^{\otimes^{m\!-\!j\!-\!1}}_{\text{odd}}
+|1\rangle^{\otimes^{j\!-\!1}}_{\text{odd}}\sigma_{x}\sigma_{z}^{\omega_{h,m}}|\phi\rangle_{j}|1\rangle^{\otimes^{m\!-\!j\!-\!1}}_{\text{even}}\\
+|1\rangle^{\otimes^{j\!-\!1}}_{\text{even}}\sigma_{z}^{\omega_{h,m}}R_{z}(\theta_{s})|\phi\rangle_{j}|1\rangle^{\otimes^{m\!-\!j\!-\!1}}_{\text{even}}
+|1\rangle^{\otimes^{j\!-\!1}}_{\text{odd}}\sigma_{z}^{\omega_{h,m}}R_{z}(\theta_{s})|\phi\rangle_{j}|1\rangle^{\otimes^{m\!-\!j\!-\!1}}_{\text{odd}},
\end{split}
\end{equation}
Therefore, any one $\textbf{\emph{r}}_{j}$ of the receivers and the other $m\!-\!1$ receivers can perform  respectively the Pauli operator $\sigma_{z}^{\omega_{h,m}}\sigma_{x}$ and $\sigma_{z}^{\omega_{h,m}}$, which will arrive at
\begin{small}
\begin{equation}
|1\rangle^{\otimes^{j\!-\!1}}_{\text{even}}|\phi\rangle_{j}|1\rangle^{\otimes^{m\!-\!j\!-\!1}}_{\text{odd}}
+|1\rangle^{\otimes^{j\!-\!1}}_{\text{odd}}|\phi\rangle_{j}|1\rangle^{\otimes^{m\!-\!j\!-\!1}}_{\text{even}}
+|1\rangle^{\otimes^{j\!-\!1}}_{\text{even}}\sigma_{x}R_{z}(\theta_{s})|\phi\rangle_{j}|1\rangle^{\otimes^{m\!-\!j\!-\!1}}_{\text{even}}
+|1\rangle^{\otimes^{j\!-\!1}}_{\text{odd}}\sigma_{x}R_{z}(\theta_{s})|\phi\rangle_{j}|1\rangle^{\otimes^{m\!-\!j\!-\!1}}_{\text{odd}},
\end{equation}
\end{small}which is equal to
$|\Phi_{h,s}^{\prime}\rangle_{r}^{m}$, which can be contained by the same way above  based on the following  equation
\begin{equation}
|a\rangle|\phi\rangle
+|(a\!+\!1)\,\text{mod}\, 2\rangle\sigma_{x}R_{z}(\theta_{s})|\phi\rangle=|\phi\rangle|a\rangle
+\sigma_{x}R_{z}(\theta_{s})|\phi\rangle|(a\!+\!1)\,\text{mod}\, 2\rangle,\,a=\{0,1\}.
\end{equation}

\section{References}

\bibliographystyle{spphys}       
\bibliography{mybibfile}

\end{document}